\documentstyle[twoside,fleqn,npb,epsfig]{article}
\newcommand \Pomeron {I\!\!P}

\newcommand{\AmS}{{\protect\the\textfont2
  A\kern-.1667em\lower.5ex\hbox{M}\kern-.125emS}}
\title{Future Small $x$ Physics
  with $ep$ and $eA$ Colliders}

\author{L. Frankfurt\address{School of Physics and Astronomy,
\\ Tel Aviv University,
         69978 Tel Aviv, Israel}
and M. Strikman\address{Deutsches Elektronen Synchrotron, DESY, \\
Notkestrasse 85, 22603 Hamburg, Germany}
        \thanks{
On leave of
absence from Department of Physics, Pennsylvania
State University, University Park, PA 16802, USA}}

\begin{document}

\begin{abstract}
The interaction of spatially small dipoles
with nucleons, nuclei is
calculated in the DGLAP approximation
 at the top of HERA
 energies   and found to be close 
to
the $S$-channel unitarity limit in the  case of the color 
octet dipoles. The DGLAP analyses of the 
current diffractive data appear to support
this conclusion as they indicate a $\sim 30-40\%$ probability 
of the gluon induced diffraction for
 $Q^2\sim 4\,$GeV$^2$. The  need for 
the high-precision measurements of the $t$-dependence of 
inclusive and exclusive diffraction 
for pinpointing higher twist effects in the gluon sector is emphasized.
The $eA$ collisions at HERA would 
provide a strong amplification of the 
gluon densities allowing to reach deep into the 
regime of nonlinear QCD evolution.
Connection between the leading twist nuclear shadowing
and leading twist diffraction in $ep$ scattering is explained. The 
presented
model independent results for the nuclear shadowing for light nuclei 
indicate much larger shadowing for the gluon sector
than for the sea quark sector.
It is argued that HERA in $eA$ mode would
be able to discover a number of new phenomena including
 large gluon shadowing,
large nonlinearities in parton evolution, small $x$ color
transparency in the vector meson production followed by 
color opacity at $x\le .005$, large probability 
of inclusive diffraction.
Implications
for the nucleus-nucleus collisions at LHC are discussed as well.

\end{abstract}

\maketitle

 \section{INTRODUCTION}  
   Fascination with small     $x$ - physics can be traced back to the old   
questions - why is the 
cross section of  nucleon-nucleon interactions at    
high energies   large   and slowly growing with   $s$? Can 
the interaction  of a heavy onium state with hadrons
remain weak even at very high energies? 
Is it possible to exploit
 the increase of the parton distributions
at small $x$ to produce superdense  hadron matter in the laboratory?   

  A unique advantage of DIS for studying  dynamics of strong   
interactions is the possibility    {\it to tune} the resolution scale.   
Nuclei add another   dimension - possibility {\it to vary}  the 
thickness of the target. Hence it is possible to 
investigate the transition from  soft nonperturbative dynamics to    
perturbative dynamics: up to what virtualities (or down to which
distances) is   nonperturbative QCD important? In addition,  how  does
the transition depend on the incident energy?  

The fundamental questions one can address in DIS at small $x$
are: How large is the cross section of interaction of  small 
``hadrons''  (oniums)  with nucleons/nuclei at high energies? 
Is a kind  of the Froissart bound for DIS close to the
HERA kinematics for the gluon induced processes?,
Would nuclei be transparent to a high-energy 
$J/\psi, \Upsilon$? How far down in $ x$ it is necessary to go
for the observation of a break down of the linear DGLAP evolution 
equations?
What are
 implications of nonlinear QCD evolution for the physics of $AA$ collisions?

In this talk we first focus on $ep$ scattering. We discuss
the interaction of spatially small color singlet clusters with 
hadrons at high energies.
The $S$-channel unitarity constraint on this 
interaction implies  that, in the 
case of color octet dipoles interaction may already be 
close to the unitarity limit already at higher HERA energies,
implying a large probability of
diffraction when the hard process is induced by 
the interaction with 
a gluon. The use of the QCD factorization 
theorem  \cite{Collins} 
for inclusive diffraction allows the  extraction of  this
probability from the current analyses of the HERA data. 
The probability is indeed large 
$\sim 30-40\%$ in a wide range of $Q^2$ which is 
 much larger than the measured probability of the quark 
induced diffraction. We found that breakdown of the DGLAP 
approximation for the gluon channel (higher twist effects 
becoming comparable to the leading twist effects)
is related to large parton densities at low Q scale rather than to the
BFKL type dynamics \cite{BFKL}.
It should be accompanied by large fluctuations of gluon densities. 
Further directions of studying the gluon dynamics at small $x$ at HERA are 
outlined.

 In the second part of the talk, we explain how the study of
electron-nucleus collisions 
at HERA can add  to the understanding small
 $x$ dynamics. We emphasize that a number of expected
 phenomena follow from the basic features of 
the small $x$ dynamics. In particular,
the application of the $S$-channel unitarity shows that
 nonlinear effects in the gluon channel
already are large.  The Gribov black body limit
for the cross section of DIS \cite{Gribov}
is found to be a good guess for  gluon interactions in 
the HERA kinematics.
Combining the Gribov theory of shadowing and the QCD 
factorization theorem for diffraction allows the  derivation of the 
model independent expressions for the leading twist quark and gluon
shadowing for the scattering off light nuclei. An enhancement of  
gluon shadowing, compared to  quark shadowing, is predicted.
It is concluded that nuclear matter at HERA 
energies should be opaque to the small size dipoles, leading to break down
of color transparency regime
at small $x$ ($x\le 0.005$ for $J/\psi$ photoproduction).
Large gluon shadowing would lead to significant changes of the 
first stage of the heavy ion collisions at LHC.
Really account of nuclear shadowing within the conventional 
convolution formulae leads to a strong (a factor $\sim 10$) 
suppression of the yields of minijets.

\section{ELECTRON-PROTON  PHYSICS}
\subsection{The interaction of small color singlets with hadrons} 
 At small $x$  the average longitudinal distances   
(Ioffe distances) in the correlator of the e.m. currents which determine
$F_{2p}(x,Q^2)$ are    $ l_{coh}\sim {1\over 2m_Nx}$. 
At HERA   $l_{coh}$ can reach  $10^3$ fm. 
Hence in the target rest frame the small $x$ processes 
occur in three steps. Initially 
the  $\gamma^*$ transforms into $ q\bar q$ pair.  Next, the 
$\bar 3 3$ color dipole  of transverse size     $b$ interacts   
with the target. This cross section   
rapidly grows with increase of energy  \cite{BBFS,FRS}: 
\begin{equation} 
\sigma_{q\bar q,N}^{inel}(E_{inc})={\pi^2\over   
 3}b^2\alpha_s(Q^2)xG_N(x,Q^2\simeq{\lambda\over b^2}),   
\label{sigq}
\end{equation}
where    $\lambda(x \approx 10^{-3})\approx  9,   
x={Q^2\over 2m_NE_{inc}}$.    
 
 For color octet dipole 
the cross section is 
enhanced 
by the ratio of Casimir operators of color group $SU(3)_c$ - ~
 $C_F(8)/C_F(3)= 9/4$   \cite{AFS,FRS}:  
 \begin{equation}   
\sigma_{g~g,N}^{inel}(E_{inc})={3\pi^2\over   
 4}b^2\alpha_s(Q^2)xG_N(x,Q^2\simeq{\lambda\over b^2}),   
\label{sigg}
\end{equation}  
These equations are  another form of the QCD evolution equations.   
For the fit to 
the energy dependence  
\footnote{Such a fit is useful in the practical applications
although QCD predicts a behavior $\propto a +b\ln{1\over x}
+ c\ln^2{1\over x}$ for the HERA energy range
since radiation of $\propto$ 1-2 hard gluons is possible in the multiRegge 
kinematics at HERA.} 
~
$\sigma(s,Q^2)\propto s^{(\alpha(Q^2)-1)}$, one finds
that due to increase of the steepness
of the gluon density with virtuality, $\alpha(Q^2)$
gradually increases with $Q^2$:

\begin{equation}
\alpha(4GeV^2) \approx 0.2, \alpha(40 GeV^2)\approx 0.4. 
\label{power}
\end{equation}

\subsection{Pomeron: one, two, ... too many?}      
\noindent
{\it 
Satan: ``You were selling spoiled sturgeon fish.''\\
Merchant: ``No - sturgeon was of regular quality''\\
Satan: 
``Sturgeon could only be the prime quality 
there is no other one.''

\hspace{1cm} ``Master and Margaret'' M.Bulgakov}

Eqs.\ref{sigq},\ref{sigg},\ref{power} clearly demonstrate a
qualitative difference of the strong interaction dynamics in DIS 
at small $x$ from the  soft 
physics of the vacuum exchange. This object 
was derived by V.Gribov within the
high-energy Regge theory as the rightmost singularity
in the complex angular momentum plane. It was  named Pomeron by 
M.Gell-Mann after I.Pomeranchuk who was the first to 
predict the total cross sections of particle(antiparticle)-
hadron scattering to be equal at asymptotic energies. The key
feature of the Regge trajectories is their {\it universality}
- independence of the trajectories on the process. 
Clearly the behavior of the cross sections, in the kinematics where  
DGLAP  describes the $Q^2$ evolution of the parton
densities, violates such universality in a gross way.

It is sometimes suggested that one can 
save the idea of the Pomeron in DIS processes by 
introducing two Pomeron trajectories with different 
intercepts. Clearly a smooth change of the effective power with $Q^2$
indicates that this hypothesis is not equivalent to
eq.\ref{sigq}.

Another  manifestation of the breakdown of the universality is the
change of the pattern of the $t$-dependence of the quasi two-body
processes expected in pQCD  \cite{BFGMS}.
Let us consider for example the process $\gamma^*+p\to V +p$
where V is a vector meson.

In the soft regime the $t$ dependence of the differential cross
section changes with $s$ as
\begin{eqnarray}
{d \sigma \over dt}=f(s,0)\phi(t)
 \left({s\over s_0}\right)^{2\alpha_{\Pomeron}-2}\simeq 
\nonumber \\
 f(s,0)\phi(t)\exp({2\alpha^{\prime}t\ln(s/s_0) }.
\label{slope}
\end{eqnarray} 
In the impact parameter space this amounts
to an increase of the radius of the projectile hadron with 
increase of the incident energy. This can be understood as due to
multiperipheral structure of the soft process - fast partons of the projectile
decay into slower partons with a random change of the position in 
the impact parameter space of $b \sim {1\over k_t}$ where 
$k_t$ is the soft scale (transverse momentum of the emitted
parton). The resulting random walk in the impact parameter space
leads to a growth of $\left<b^2\right>$ 
proportional to the number of steps, that is to the length
of the rapidity  interval. Hence the size grows $\propto 
\Delta y$, leading to the shrinkage of the cone given 
by eq.\ref{slope}\,  \cite{Gribovalpha}. 

In the hard regime the parton (gluon) emissions are governed by the
hard scale (Fig.1). Hence the random walk 
produces much smaller increase of  $\left<b^2\right>$:
$\alpha_{hard} \ll \alpha_{soft}$. 

Therefore with an increase of $Q^2$ one expects three
phenomena to happen  \cite{BFGMS}: (i) a decrease of
the contribution of the upper end of the ladder to 
the slope, B ($f(t)=\exp(Bt)$) 
which should tend 
to the limiting value  determined by
the two-gluon form factor of the nucleon - experimentally
$B_{2g}\approx 4$ GeV$^{-2}$. (ii) Gradual disappearance of 
the energy dependence of the slope.
(iii) A fast increase of the cross section with energy 
$\sim \left|x G(x,Q^2)\right|^2 \propto {Q\over x^{0.4}}$.
 However this pattern can be sustained only  until the
transverse momenta in the perturbative ladder degrade to the level
corresponding to the soft physics. At this point the further steps 
in the ladder will be soft and will lead to increase of the slope
(see the sketch Fig.2) and lead to
the values of $\alpha^{\prime}$ which are likely to approach, at
very high energies, values similar  to those for the soft physics.

\begin{figure}
    \begin{center} 
        \leavevmode 
        \epsfxsize=0.90\hsize 
        \epsfbox{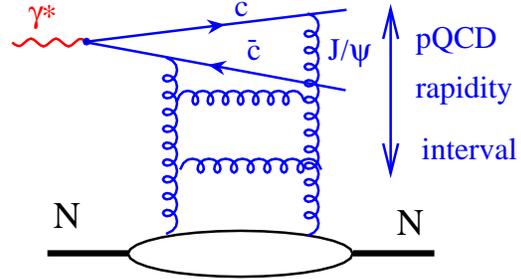} 
    \end{center} 
\vspace{-0.5cm}
\caption{Electroproduction of $J/\psi$.
The rapidity
interval occupied by hard exchanges
}
\label{psif}
\end{figure}

\begin{figure}
    \begin{center} 
        \leavevmode 
        \epsfxsize=0.90\hsize 
        \epsfbox{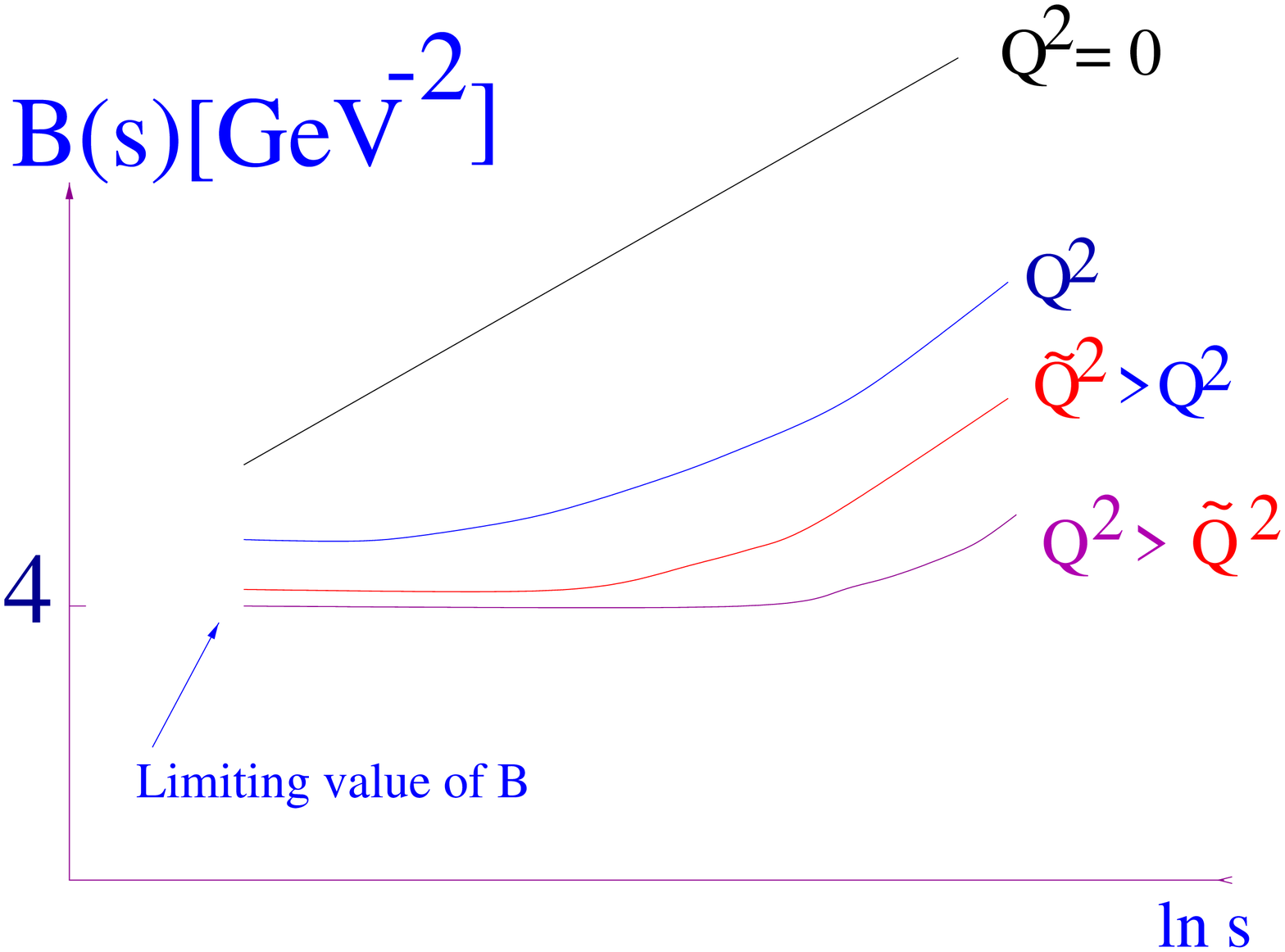} 
    \end{center} 
\vspace{-0.5cm}
\caption{Sketch of the $Q^2$ and energy dependence of the
slope of the vector meson production}
\label{sketch}
\end{figure}

Obviously the discussed pattern is qualitatively 
different from  the soft Pomeron type physics.
Hence, one should be extremely careful with 
the use of the word ``Pomeron'' in 
the DIS regime.

Current HERA data confirm the prediction of the 
convergence of the slope of  $\rho$-meson production
at large $Q^2$ to that for  $J/\psi$ (section 2.5). However
the HERA data are not sufficiently accurate so far to
determine $\alpha^{\prime}$ 
for DIS processes in the HERA energy range 
and the change of the slope between the HERA range and the fixed 
target experiment range is still a matter of debates  \cite{Levy,H1}.
Nevertheless one can obtain clear 
indications of nonuniversality of $\alpha^{\prime}$ 
by combining HERA and SLAC linac data. Indeed,
assuming for a moment
that $\alpha^{\prime}$  is universal we would find that 
$B\approx$ 4 GeV$^{-2}$ for 
$W=200$ GeV as measured at HERA would lead to
$B\approx$ 1 GeV$^{-2}$ at $W \sim$ 6 GeV, while
the SLAC linac data at that energy correspond to $B\approx 3\,$GeV$^{-2}$.

Hence we conclude that  an accurate measurement of the energy
dependence of the $t$-dependence 
would provide a unique way to determine the 
transition from soft to hard physics and 
provide a direct evidence 
for non-Regge type behavior of the 
interaction in the vacuum channel in DIS.

\subsection{Can cross sections grow rapidly  forever?}

A fast increase of the inelastic cross section of the small dipole -nucleon 
interaction cannot continue indefinitely. 
The black body limit:
$\sigma_{diff}\le \sigma_{tot}/2$    
indicates that this is impossible.    

{\it    Nucleon case   }

The analysis is much simpler than in the case of    
soft interactions since the radius of the system
practically  does not grow with energy 
(see discussion in the previous section and Fig.6).

It follows from the optic theorem that
 \cite{FKS}:
\begin{equation}   
 \sigma_{el}=(1+\eta^2){\sigma_{tot}^2\over 16\pi B_{2g}},
\label{optic}
 \end{equation}  
where $\eta$  is  the ratio of real and imaginary parts of the amplitude.
Using the Gribov-Migdal relation for $\eta$:
\begin{equation}
\eta={\pi\over 2} {d \ln \sigma_{tot}\over d\ln s}
\label{eta}
\end{equation}
and $\eta \ge 0.4 $ for $Q^2\sim 10$ GeV$^2$ from eqs.\ref{sigq},
\ref{eta}
we find a kind of the Froissart bound:
\begin{equation}
\sigma_{tot}\le {8\pi B_{2g} \over 1+\eta^2}\approx 35 mb,
\end{equation}   
for $Q^2=10$ GeV$^2$ and therefore  $\sigma_{inel}\le 18 mb$. 
The limit becomes tighter with increase of $Q^2$ due 
to increase of 
$\eta$ with $Q^2$.

To illustrate numerically to what extent this limit is relevant for
dynamics at HERA and beyond let us consider
$Q^2\sim 10$ GeV$^2$  and $x\sim 10^{-4}$  
and take $xG_N\sim 20 $ and $\alpha_S=.25$. Clearly such 
an estimate carries substantial uncertainties,
in particular because eqs.(\ref{sigq},\ref{sigg})
were derived in the leading $\alpha_s\log Q^2$ approximation.
The cross section of interaction of the color
triplet dipoles with nucleon is
still much smaller than the discussed limit: $\sigma_{inel} \sim 7$ mb.
Using eq.(\ref{optic}) we also get
$ {\sigma_{el}\over \sigma_{inel}} \approx 0.1$.
At the same time for the color  octet dipole eq.(\ref{sigg}) gives
$\sigma_{inel}(x=10^{-4}, Q^2=10$GeV$^2)\approx$ 14 mb. 
This is pretty close to the limit and  corresponds to:    
$\sigma_{el}/\sigma_{tot}\sim 1/3$.   
This indicates that decomposition of cross section over  twists
in this situation becomes unreliable, also one should expect
a  large probability of diffraction in  the gluon induced processes.  
Applying the  AGK rules  \cite{AGK} for the diagrams corresponding
to the  higher twist terms (like attachment of four gluons to the dipole)
we find that they lead to final states with double, triple,
 ... multiplicity. This  
provides an  evidence for the large fluctuations of the parton
densities in the regime close to the unitarity limit
\footnote{It is worth noting that there is no direct relation between
the higher 
twist effects in diffraction and in the total cross of DIS.
In the diffraction case we could use the optic theorem to single out
the leading term in the scattering of small clusters. In the case of
the total cross section several competing terms enter with opposite signs,
cf.   \cite{Bartels}.}

It is worth emphasizing that a large value of $xG_N(x,Q^2)$, which 
results in a breakdown of the linear evolution equations,
was generated within DGLAP predominantly due to
$log Q^2$ evolution rather than solely due to $\ln 1/x$ effects which would be
the BFKL approximation. 
Qualitatively the large gluon density is due to 
the presence of large number of
gluons already on a low resolution scale (gluon carry about the same 
momentum as quarks at this scale) and they generate a lot of gluons 
in the evolution.
(At HERA kinematics $xG(x,Q^2)\approx 20$ but only 1-3 gluons are due
to multiRegge kinematics).
When energies approach the unitarity limit the increase of $\sigma(b,x)$
should slow down to a rate which may be rather similar to the case of
soft physics. One may expect that in this limit a soft regime will be
reached with a genuine universal Pomeron.
   
{\it Nucleus case.}
In the case of a sufficiently heavy nucleus the inelastic cross section cannot
exceed the geometric size of the nucleus:
\begin{equation}
\sigma^{inel}_{q\bar q-A}(b,s)=   
{\pi^2\over 3}b^2\alpha_sxG_A(x,Q^2=\lambda/b^2)   
<\pi R_A^2
\end{equation}
Hence $xG_A(x,Q^2=10)/A\le 120 A^{-1/3}$. 
For central impact parameters the inequality is a factor of 1.5    
times stronger:   
\begin{equation}
xG_A^{cent.imp.par.}(x,Q^2=10)/A\le 80 A^{-1/3}.
\end{equation}   

 For the interactions  induced by the color octet 
dipole the constraint is a factor of {9/4} stronger:    
\begin{equation}
   xG_A^{centr.imp.par.}(x,Q^2=10)/A\le 35 A^{-1/3},
\end{equation}
which for $A=200$ is substantially smaller
than    $xG_N(10^{-4},10)\sim 20$.

Thus we conclude that at the HERA   
kinematics  quark induced interactions do
not  reach the unitarity limit 
but perhaps 
the gluon induced interactions are
close to it 
(estimates at lower Q are more uncertain but the trend is
that the
approach to the 
unitarity limit should occur earlier for smaller Q).  
 If the gluon shadowing were small,  the quark induced   
interaction at small impact parameters for heavy   
nuclei would be close to the unitarity limit for 
$Q^2\sim 10$ GeV$^2$   
leading to large nonlinear effects 
in the QCD evolution for quark sector. 

 In the case of heavy nuclei the unitarity constraint
 definitely points  to a    
modification of the dynamics for the gluon induced  
interaction even if the gluon shadowing is large.

Extension of the $x$-range by two orders of magnitude 
at TESLA-HERA collider would correspond
to an increase of the gluon densities by a factor of 3 
for $Q^2=10$ GeV$^2$. It  will 
definitely bring quark interactions at this  scale into the region
where DGLAP will break down. For the gluon-induced interactions
it would allow the exploration of a non-DGLAP hard dynamics
over two orders of magnitude in $x$ 
in the kinematics where $\alpha_s$ is small while the fluctuations of
parton densities are large.

\subsection{Inclusive diffraction}
We discussed above that a large 
probability  of diffraction  in DIS would 
be a clear signal for the onset of a new nonlinear dynamics.
Since this characteristic of DIS can be directly measured, it
provides a more direct probe than analysis of inclusive structure
functions which depend on the nonperturbative
boundary conditions which are not 
constrained by the theory strongly enough.

The current situation with diffraction at HERA is already very intriguing.  
One expects, based on the QCD factorization theorem for diffractive 
processes \cite{Collins}, that the diffractive
structure functions  $f_j^D(\beta,Q^2,x_{\Pomeron},t)$    satisfy
DGLAP evolution equations. Here $\beta=x/x_{\Pomeron}$.
The data seem to be consistent with an early scaling for $Q^2\ge 4$
 GeV$^2$
for most of $\beta $ range and a large probability of diffraction -
$P_q\approx 10\%$ for $\gamma^*p$ scattering poses 
rather strong constrains on the dynamics.

  This experimental observation is   highly non trivial.
Indeed, if the DGLAP evolution for the fragmentation  would  be valid 
for    $Q\ge Q_0=0.7~$GeV where quark and gluon parton densities tend 
to zero at $x\to 0$ (the GRV scenario) the rapidity gap probability in
DIS would be $\ll  P_q\approx  0.1$    
observed at HERA.   (Naively, if one assumes
factorization for the inclusive cross sections, the 
assumption of an early factorization for the leading baryon
production should be reasonable as well. 
Indeed, the overlapping integral for the leading partons
(present at the scale $Q_0$ )
with the emitted partons to form a 
low  $p_t$     leading nucleon should be very small
since the partons which are generated in the evolution have
$p_t\ge Q_0$ and rather small light-cone fractions  
).

Another interesting feature of the current data (shared by the 
diffractive data from the proton colliders) is a
very important role of the 
gluons in the diffractive events. In the parlor of diffractive community -
the ``Pomeron'' is predominantly build of gluons. To quantify this
statement it is convenient to define the probability
of diffraction for the action of the hard probe which couples to a
parton $j$:
\begin{equation}
P_j(x,Q^2)=   
{\int f_j^D({x\over x_{\Pomeron}},Q^2,x_{\Pomeron},t)dt dx_{\Pomeron}\over   
f_j(x,Q^2)}.
\end{equation}
 If the interaction in the gluon sector  at small $x$ reaches   
strengths close to   
the unitarity limit we should expect that   $P_g$ is
 rather close to 1/2  and much larger than   $P_q$.   
 Using eq.\ref{sigg} for    
    $\sigma_{color ~ octet~dipole -N}^{inel}$  we calculate
     $P_g(x=10^{-4},Q^2=10$ GeV$^2)  \approx 1/3$.    
 Indeed, using the global fit of Alvero, Collins and Whitmore \cite{ACW}
 we find (Fig.3):
  $P_g \gg P_q$, and $P_g(x\le 10^{-3}) \approx 0.4 (0.3)~$ for 
$Q^2=4(10)$~GeV$^2$!
 \begin{figure}
\vspace*{-0.8cm}
    \begin{center} 
        \leavevmode 
        \epsfxsize=1.0\hsize 
        \epsfbox{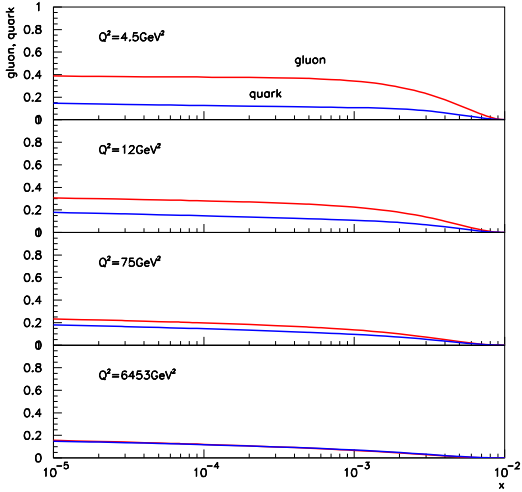} 
    \end{center} 
\vspace{-1cm}
\caption{Probabilities of rapidity gap events for quark and gluon 
induced diffraction calculated using the best fit of  \cite{ACW}.}  
\label{sigx}
\end{figure}    

It is also interesting  that diffraction when parameterized in terms of
the effective Pomeron trajectory corresponds to $\alpha_{\Pomeron}(t=0)
=1.15-1.2$ which is somewhat larger than what is usually
observed in the soft domain.

A word of caution: the direct measurements of   $f_g^D$ are    
available for a limited   
range of     $x_{\Pomeron}$ and for rather large       
$Q^2$. ``Backward'' evolution neglects higher twist effects which may   
slow down this evolution and reduce $P_g$ at    $Q\sim Q_0$. 
However backward evolution should be a small effect for    
$Q^2 \sim 15$ GeV$^2$  where    $P_g \sim 0.3$.   
So we conclude that the data confirm the expectation that 
interactions in the gluon sector are strong. They are not 
corresponding to the black limit since $P_g$ is substantially smaller 
than 0.5. Also, 
the measured slope of diffraction suggests that configurations involved
in the total cross section of  diffraction have a rather large size
while the slope of the cross section
of diffraction in the gluon induced processes was not measured 
so far.

\begin{figure}
\vspace*{-0.7cm}
    \begin{center} 
        \leavevmode 
        \epsfxsize=0.90\hsize 
        \epsfbox{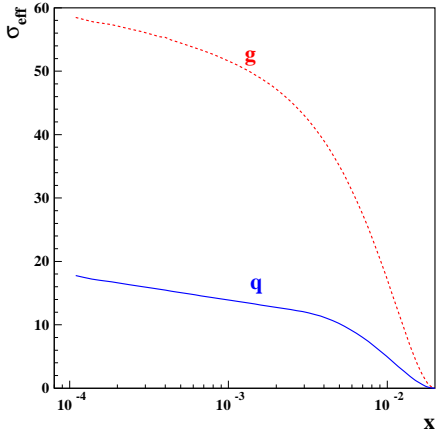} 
    \end{center}
\vspace*{-1.6cm} 
\caption{The $x$-dependence of $\sigma_{eff}$ for
quark and gluon induced diffraction.}
\label{sigdif}
\end{figure}

Due to the 
large distances involved in the diffraction we can think of diffraction
as ``elastic'' scattering of $q\bar q$ or $gg$ configurations of the target
(if we define the process at a given $Q$ scale, changing the scale 
leads to a mixing of these configuration, and the addition
of new intermediate states like $q\bar q g$). Hence it
is instructive to treat diffraction in a complementary   
$S$-channel picture -the eigen states    
of the scattering matrix  \cite{eigen}.   
   Use of the optic theorem  leads to
\begin{equation}
\sigma_{eff}(x,Q^2)\equiv {16\pi d\sigma^{dif}/dt_{\left.\right|   
 t=0 }   
\over \sigma_{tot}}
\end{equation} 
and allows to extract the average     
cross section for the configurations which contribute to   
quark and gluon induced diffraction:
  $\sigma_{eff}^{q(g)}(x,Q)$. 
We find
$\sigma_{eff}^{gluons}(x=10^{-4},Q=4)\sim 30 mb$ and even   
larger   value of 
   $\sigma_{eff}^{gluons}(x=10^{-4},Q=2)\sim 50 mb$ which
  indicates    
presence of superstrong       
interactions in the gluon sector (Fig.\ref{sigdif}).
These strong interactions should be   
due to some hard  dynamics of strong gluon fields   
since the $\gamma+p\to J/\psi+p$ data suggest that pQCD   
for these Q occupies most of the rapidity interval.

  This argument for superstrong gluon interactions at small   $x$   
is complementary to the unitarity argument given    
above.

Several avenues of studies are possible in $ep$ scattering
to investigate relative role of 
soft and hard physics in the gluon induced diffraction:
detailed    investigation of charm diffraction, diffractive
dijet production, energy dependence of the $t$-slope.   
The study of scattering off nuclei - nuclear shadowing, diffraction
off nuclei  would be able to provide direct answers to these questions.   
   
To summarize, studies of diffraction at HERA 
appear to confirm the conclusion
of section 2.3 that large  nonlinear effects could be taking    
place in $ep$ scattering for      $Q^2 \le 4-10$ GeV$^2$  in the gluon    
sector - the main problem  is the  lack of effective
``gluonometers'' 
(especially in diffraction) at this resolution scale.   

\begin{figure*}  
\vspace*{-0.6cm}
     \begin{center} 
        \leavevmode 
        \epsfxsize=0.90\hsize 
        \epsfbox{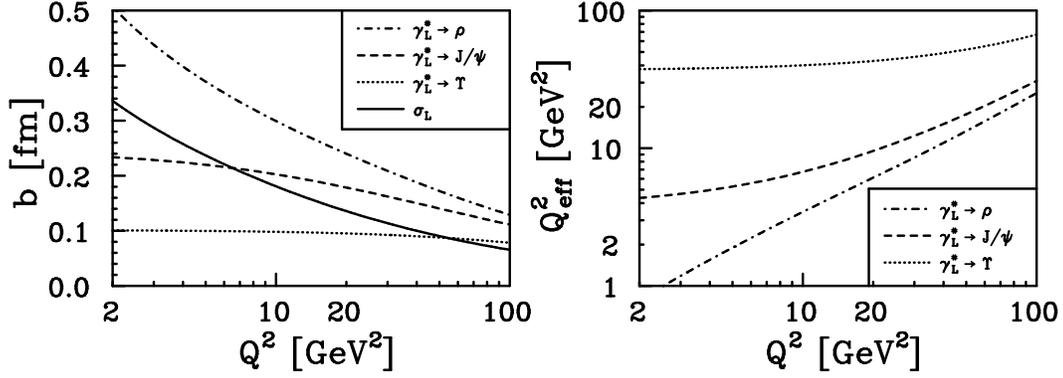}
    \end{center}
\vspace*{-1.0cm}  
\caption{The dependence of average $b$ and effective $Q^2$ of $Q^2$
for production of vector mesons  \cite{FKS}} 
\label{bsize}
\end{figure*} 
\subsection{Vector meson production - a quest for a low Q gluonometer}
Over the last few years a lot of experimental and theoretical 
studies of the exclusive vector meson production in DIS were performed.
A proof of the QCD factorization theorem for exclusive
meson production by longitudinally polarized photons 
 \cite{CFS} has demonstrated that 
for large Q 
these processes may be calculated as rigorously as the leading
 twist processes. In the small $x$ limit
the amplitude of the process - 
$A(\gamma^*_L+p\to V+p)$ 
 can  be written as a convolution of the light-cone
wave function of the photon
  $\Psi_{\gamma^* \rightarrow |q\bar q\rangle}$,
the scattering amplitude of the hadron state,  $~\sigma(q\bar q T)$,
and the wave function of the vector meson,  $\psi_{V}$.
$\sigma(b,s)$ is expressed
 through the skewed gluon  
density of the target \footnote{At small
$x$ and large $Q^2$ the skewed parton
distributions are calculable in QCD as a solution of the QCD
evolution equation with the skewed kernel and the conventional
parton distribution at low scale as the nonperturbative input.
 \cite{FFGS}}.
In  impact parameter space (suppressing the z-dependences)
\begin{equation}
A= \int d^2b \psi_{\gamma*,L}(b)\sigma(b,s)\psi_V(b).
\label{over}
\end{equation}

The parameter free  leading twist answer  for the 
process is
\begin{eqnarray}
{d\sigma^L_{\gamma^*N\rightarrow VN}\over dt}\left|_{t=0}\right. =
12\pi^3\Gamma_{V \rightarrow e^{+}e^-} M_{V} \cdot\nonumber \\
{\alpha_s^2(Q)\eta^2_V \left|\left(1 + i{\pi\over2}{d \over 
d\ln x}\right)xG_T^{skewed} (x,Q^2)\right|^2 \over \alpha_{EM}Q^6N_c^2}
\label{psi}
\end{eqnarray}
Here,
 $\Gamma_{V \rightarrow e^{+}e^-}$ is the decay width of
  $V\to e^+e^-$;
$\eta_V\equiv {1\over 2}{\int{dz\,d^2k_t\over z(1-z)}\,\Phi_V(z,k_t)\over
\int dz\,d^2k_t\,\Phi_V(z,k_t)}\rightarrow 3$ for $Q^2\rightarrow \infty$.

 In the leading twist  b=0 in $\psi_V(b)$. Finite  b effects in the 
meson wave function in the overlap integral (\ref{over})
appear to be one of the major sources of  higher twist effects at small $x$.
They strongly reduce the cross section at intermediate $Q^2$.
However the transverse sizes essential in the amplitude may  remain small,
down to rather small $Q^2$ (Fig.\ref{bsize}).
Hence it is likely that the study of $\rho,\phi$-meson production
at intermediate $Q^2 \sim 5-10$ GeV$^2$ will
 allow the study of the gluon densities at $Q^2$ down to 2-3 GeV$^2$.
The best way to check the quality of $\rho,\phi$-mesons as low-Q 
gluonometers
would be to study color transparency in the 
production of these mesons
at intermediate $Q^2$, see discussion in section 3.7.1.

\begin{figure}
\vspace*{-1.5cm}
\begin{center}
\leavevmode
       \leavevmode 
        \epsfxsize=.9\hsize 
        \epsfbox{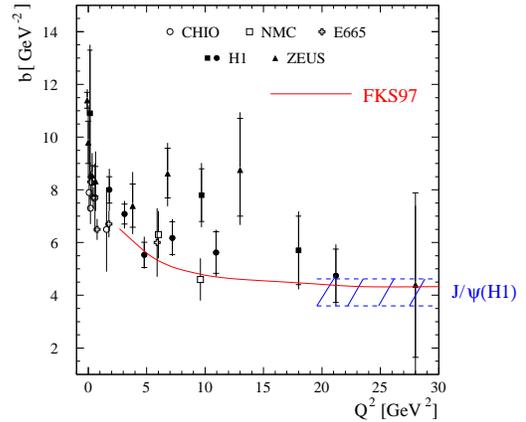}
\end{center}
\vspace*{-1.3cm}
\caption{Comparison the DESY data \cite{H1rho} for the slope of the $\rho$
meson production with the calculation of  \cite{FKS}.}
\label{bfig}
\end{figure}

Experiments at HERA have confirmed a number
of the QCD predictions \cite{BFGMS} for the vector meson production:
(i) The rate of the increase with energy for $\rho$ production
for $Q^2 \ge 20$ GeV$^2$ and for $J/\psi $ production
is $\left|xG_N(x,Q^2_{eff})\right|^2 \propto W^{0.8}$
for  $Q^2_{eff} \sim 4$ GeV$^2$. (Note that the soft physics expectation is  
$\sim W^{0.32}$).
The models which take into account 
higher twist effects based on eq.(\ref{over})
 \cite{FKS,GV} are able to
 to describe the HERA data at lower $Q^2$. The inclusion
of the quark exchange allows to reproduce the magnitude of the
fixed target data as well  \cite{GV}.
(ii) absolute cross section of $\Upsilon $ production 
is reasonably reproduced  \cite{FMDS}. 
(iii) The extensive data of H1  \cite{H1psi} on the energy and $Q^2$
dependence of $J/\psi$ production agree with predictions of  \cite{FKS}
and confirm need for large higher twist effects. The model of
Ryskin  \cite{Ryskin} which neglects these effects leads to a factor
$\sim 4-5 $ faster decrease of the cross section with $Q^2$,
(iv) the dominance of  $\sigma_L$ over  $\sigma_T$
  at $Q^2\gg m_V^2$ is confirmed (iv) The $\phi/\rho$ ratio has increased
a factor $\sim 4 $ as compared to the value at $Q\sim 0$
and reached the value close to the QCD prediction of 1.2*(2/9) 
(where 2/9 is the SU(3) result), similar trend is observed for the
$J/\psi/\rho$ ratio.
(v) Convergence of the $t$-slopes of $\rho$ meson production at large
$Q^2$ and the $J/\psi $ slope (which weakly depends on $Q^2$)
 has been
observed at HERA. The $Q^2$ dependence of $B$
can be described semiquantitatively as a result of the finite $b$
in the vector meson wave function, see Fig.\ref{bfig}.
(vi) The data are consistent with a slower shrinkage of
diffractive cone for these processes than for soft phenomena, see
discussion in section 2.2.

\subsection{BFKL dynamics -
small dipole -  small dipole    scattering}
The cross section of scattering 
of two small dipoles was calculated in the leading
log 1/$x$ approximation neglecting  $\ln Q^2$ evolution 
by BFKL  \cite{BFKL}.

Calculation of next to  leading $\log 1/x$ effects were reported 
a year ago which demonstrated
a large reduction of the rate of increase with energy, as compared to
 the leading $\log 1/x$ approximation. To a large extent this reflected
the smallness of the phase space - emission of one gluon in this approximation
requires at least
about two units of rapidity (at HERA this corresponds to emission
of $\le 3 $ gluons).
A number of attempts to perform resummation
of NLO BFKL were reported recently. The current
scenario is that 
$ \sigma \propto s^{.25\pm 0.05}$ and
that  diffusion in the impact parameter space may be rather weak and
not  bring
an  onset of nonperturbative regime over a significant 
energy range  \cite{talks}.

This scenario leads to 
a slower energy dependence of the cross section than eq.\ref{sigq} 
for  the scattering of a small dipole off a large hadron.
Such a pattern seems unrealistic and reflects that  in  the
HERA range  the $\ln Q^2$ effects which are neglected in the BFKL model
are very important. Hence the quantitative comparison of the model
with the data is hardly possible at present. It would be
 more promising to study the  $p_t$ ordering
of the produced hadrons which is qualitatively different in BFKL and DGLAP
approximations.  Therefore detection of particles
over large
 rapidity interval is necessary - 
to emit three gluons between dipoles 
in BFKL kinematics  $\Delta y \ge 2$
between the jets one needs a detector
with acceptance of  $\ge 8$ units in rapidity.
Promising direction of the study of the BFKL regime is 
$\gamma^*-\gamma^*$ scattering at large $Q_1^2 \sim Q^2_2$.
In this case contribution of the 
scattering of two small dipoles is enhanced by a power of $Q^2$ as 
compared to the scattering of a small dipole on a
 large dipole \cite{gamma}.

\subsection{Main focus of the future studies}
Since the most dramatic effects are expected to be happening 
in the processes directly coupled to gluons
the major focus of the studies should be
a measurement of the quantities
more directly sensitive to the gluon dynamics
and to the onset of hard (pQCD) regime. This includes measurement of 
$\sigma_L(x,Q^2)$, 
the absolute value,  $W,Q^2,t$ dependence of diffraction for 
 the exclusive production both for $\sigma_L$ and $\sigma_T$
(vector mesons, 
$\pi \pi$ ....), for   charm, dijets, etc. 

The key issue is to establish
 nonuniversality of the vacuum exchange -
the $Q^2$ and flavor dependence of the
rate of the change of the $t$-slopes with energy,
and 
of the
energy dependence of the diffractive processes, and especially the 
difference between the hard diffraction induced 
by hard $\gamma^*$-gluon and by $\gamma^*$-quark 
scattering.

Another direction of the research would be looking for manifestations
of  BKFL-type dynamics in the hard processes spanning maximal possible
range in rapidity.

To achieve these objectives one would need to perform 
high precision 
DIS measurements both at 
lower energies down to $W \sim 10 GeV$ and beyond the 
energies of HERA - future $ep$ colliders.
Accurate measurements
of the $t$ dependence of diffractive processes requires improving the
forward acceptance of the collider detectors.
The extension of the detector rapidity acceptance range much 
further into the proton fragmentation region would be necessary to
reach the rapidity ranges of $\Delta y \sim 6-8$ which would allow
studies
of a new pattern of $p_t$ ordering important in the kinematics where
$\log {1\over x}$ effects become essential.

\section{SMALL $x$ PHYSICS IN $eA$ SCATTERING}
\subsection{High-Energy $eA$ scattering - History}
Thirty  years ago nuclei were suggested as a direct way to observe
 hadron properties of light, see review and references
 in   \cite{VDM}.

The striking effect was that 
in spite of the small 
value of  $\sigma_{\gamma N}$, the screening was observed
for $\gamma^*$ -nucleus cross sections:
 ${\sigma_{\gamma A}\over A\sigma_{\gamma N}} < 1$.
The reason is that
a virtual photon transforms to a hadronic (quark-gluon)
configuration  $\left|h\right>$ which propagates distances
$l_{coh} = {2 q_0\over Q^2+M_h^2}  \approx {1\over
2m_Nx}$~~ in ~~the ~~DIS ~~limit,
which by far exceed the nucleus size of  $2R_A$
 (large Ioffe distances). This key feature of space-time 
development of the high-energy DIS
is a backbone
of many analyses of small $x$ DIS at HERA which we discussed in section 2.

Hence there are two distinctive kinematics for DIS electron - nucleus scattering. At $x\ge 0.2$ nuclear effects in the total cross section originate from
the short-range nuclear structure, while
at small $x\ll 0.1, l_{coh}\gg 2R_A$ cross section is  practically 
not sensitive to details of the 
nuclear wave function.

\subsection{Structure functions of nuclei in the black disk limit}
In the limit 
 $l_{coh}(x) \gg 2R_A$ ( $x\ll 10^{-2}$ for A=200)
 there is   a deep connection between 
the   $\gamma^*A$
scattering and  
$R^{e^+e^-}(M^2)\equiv
\sigma(e^+e^-\to hadrons)/
\sigma(e^+e^-\to \mu^+\mu^-)$  \cite{Gribov}
\begin{eqnarray}
\sigma_{tot}(\gamma^*A)= \nonumber \\
{\alpha \over 3 \pi}
 \int_{M_0^2}^{\infty}{\sigma_{tot}(``h''-A) R^{e^+e^-}(M^2)
M^2 \over (Q^2+M^2)^2}d M^2,
\end{eqnarray}

If one makes an assumption natural for the soft
strong interaction dynamics that {\bf  ALL} hadronic
configurations $\left|h\right>$ in  $\gamma^*$
 at large enough  $\nu\equiv E_{\gamma^*}$
would  interact with a  large nucleus 
with $\sigma_{tot}(``h''-A)\sim 2\pi R_A^2$
 a gross violation of the Bjorken
scaling should take place for  $Q^2=const, \nu\to \infty$
 \cite{Gribov}:
\begin{eqnarray}
{1\over Q^2}F_{2A}(x,Q^2)
  ={\pi R_A^2\over 12\pi^2}R^{e^+e^-}\ln {x_0\over x}
\end{eqnarray}
where $x_0={1\over 2m_N R_A}$.

Since the Bjorken scaling was considered a
natural behavior of the cross sections at large $Q^2$ this prediction
of the gross violation of the Bjorken scaling was refereed by Bjorken as
the Gribov paradox. Bjorken suggested
the parton model solution of the paradox: 
the Aligned Jet Model - only a small fraction 
 $\sim 1/Q^2$
of configurations
in $\gamma^*$ which have small  $k_t$ (large transverse size)
could interact with  hadron-like cross sections \cite{BJ71}.
 The
rest of configurations would  not interact (parton model) or interact
with  in a color transparent way (pQCD of intermediate energies).

In QCD, due to a fast increase with energy
of the cross section of interaction of a  
``small color  dipole'' with a nucleon, 
the Gribov  scenario becomes ones again  a viable possibility
in the  $Q^2$ =const, $q_0 \to \infty$ limit for 
 $F_2, F_L$. In the case of gluons it leads to
 \cite{FS98}:
\begin{eqnarray}
 xG_A(x,Q^2)/Q^2= 
{\pi R_A^2 \over 8\pi^2}\ln {x_0\over x}.
\label{blg}
\end{eqnarray}
Hence 
$xG_A(10^{-4},Q^2=10)/A \sim {40 \over A^{1/3}}$ which is about 
~ 6 ~for~~ A=200 and should be compared to
the current fits which give $xG_N(10^{-4},Q^2=10)\sim 20$.
Thus eq.(\ref{blg}) implies that at HERA a strong reduction of the 
gluon densities in heavy nuclei should be observed. The interaction in the 
gluon sector in the case of heavy nuclei 
would be close to 
the black body limit  corresponding  to
 a drastic violation of the DGLAP evolution.

\subsection{Summary of the HERA 95-96 $eA$ study}
The study performed within the framework of the workshop
``Future of HERA''  \cite{report} has identified several
fundamental questions which can be addressed within the electron-nucleus
program: (i) amplification of the nonlinear effects
expected in QCD at small $x$,
(ii) the dynamics of high-energy interactions 
of small color
singlet systems with nuclei,
(iii) investigation of propagation of 
 quarks and 
gluons  through nuclear matter - energy losses, $p_t$
broadening, extra hadron production, (iv) 
connection to the heavy ion physics.

HERA would allow to use three complementary tools:
inclusive parton densities - quark and
gluon nuclear shadowing, exclusive and inclusive hard diffraction,
leading and central hadron production
in inelastic $eA$  collisions.
  Measurements with a set of nuclei with 
a luminosity  $10 pb^{-1}/A$ per nucleus would allow
high precision studies of the dependence of various characteristics 
of DIS as a function of  A at 100 times smaller $x$ than
in the fixed target experiments (which could study a small
fractions of the key  observables available for a collider).
  
  Frequent switching between 
different ions would lead to a strong reduction of 
the detector related  systematic errors 
 for  various nuclear  ratios.
  Systematic errors
 due to the radiative corrections are also small.

In particular, it would be possible to 
measure  the 
difference   the slopes  $d(F_2^A/F_2^D)/d\ln{Q^2}$
and  $dF_2^D/d\ln{Q^2}$ with accuracy $\sim 5\%$  (1pb$^{-1}$/A
per nucleus) and measure directly 
  $G_A/G_N$
 with accuracy of better than 10\%  (10-20pb$^{-1}$/A
per nucleus).  This would  allow to
observe 10\% deviations from the DGLAP predictions.

To achieve the desired luminosity it would be necessary to 
have luminosity  decreasing approximately  $\propto {1 \over A}$. 
With a reasonable capital investment this condition can be 
almost satisfied for
$A\sim 16$, while to reach this goal for heavy nuclei it appears
 necessary
to perform cooling of heavy nuclei. In principle, the cooling may be 
easier for 
heavy ions  than for protons. Its feasibility at HERA is
currently 
is under investigation \footnote{M.S. is indebted to F.Willeke and
J. Maidment
for discussions of the accelerator related  issues.}.
 
The development 
of the capabilities to accelerate nuclei would make it possible  also
to perform studies of the neutron parton densities using
the deuteron beams and the proton tagging and hence to measure the 
 nonsinglet parton densities. If the scheme for the accelerator of
the polarized deuterons could be developed 
\footnote{V.Skrinsky, private communication}
this would allow to measure simultaneously  the 
singlet and nonsinglet $g_{1N}$ which would be very beneficial
 for the spin program.

\subsection{Infinite momentum frame approaches to nuclear shadowing}

There exist two complementary approaches to 
  descriptions of the small $x$ 
physics: one is based on the use of the parton frame (fast nucleus)
another on the use of the nucleus rest frame.

The fast frame descriptions are convenient for the pQCD analyses, 
modeling  of nonlinear 
 effects. They usually assume that there is 
 no shadowing in the leading twist  at $Q_0$
 \cite{MQ,EQW,AGL,HLS,JX}.

The rest frame descriptions allows 
to visualize   connection to the  soft
physics  and to the  $ep$ diffraction  \cite{FS88,kwi,FS89,FLS,NZ,Piller,Kop,FSAGK,Barone,Orsay}. 
So far the effects of the higher twist   effects
were not treated explicitly in  these approaches.

In the  fast frame   
nucleon looks as a  {\it  ``MATRESHKA''}: the 
fast partons in nucleons  are contracted to 
a pancake of radius  $R_A$ and a small longitudinal
 size  $z=2r_N{m_N\over p_N}$
while the small $x$ partons form pancakes of 
   ${\left<x_v\right>\over x}$ times larger
size ($\left<x_v\right>\sim 0.2$ is  average $x$ of the valence
quarks). 
A nucleus looks as a collection of  {\it MATRESHKA's}
with valence quarks  not overlapping and clouds of partons of different 
nucleons  with 
$x\ll \left<x_v\right>{r_N\over R_A}$
 completely overlapping.

 One can consider small  $x$ parton density
per unit transverse area:
 ${4 \over 3}\rho_0R_AxG_N(x,Q^2),$
where  $\rho_0 \sim 0.16 fm^{-3}$ is the average nuclear density.
At central impact parameters  parton density/unit area is  1.5
times larger:
 $2\rho_0R_AxG_N(x,Q^2)$.
Hence one finds a
    large enhancement of the parton
 densities/area 
 in nuclei as compared to nucleons. For the central impact parameters:
  ${G_A/\pi R_A^2 \over {G_N/\pi r_N^2}}
\approx A^{{1\over 3}} $
${G_A\over AG_N} \approx 6_{\left|A=200\right.}$, if
$ {G_A\over AG_N}\approx 1$  

   One can study interaction   of overlapping tubes of   partons in the perturbative  strongly interacting phase 
In the simplest model of Mueller and Qiu \cite{MQ}
nonlinear effects are described by the 
fan diagrams.   The additional nonlinear term
contribution  $\delta G_A(x,Q^2)$   in the evolution equation for
 $G_A(x,Q^2)$: 
\begin{eqnarray}
Q^2 {\partial \over \partial Q^2}{ \delta x G_A(x,Q^2) \over A}=\nonumber \\
-{81 \over 16}{A^{1/3}\over Q^2r_0^2}\alpha_s^2(Q^2)\int^1_x{du\over u}\left[
uG_N(u,Q^2)\right]^2.
\end{eqnarray}
For  $A\sim 200$ this correction 
is three times larger than for the proton and
would  strongly  reduce   
 ${\partial (G_A/A)\over \partial  \ln Q^2}$ at 
 $Q^2=4$ GeV$^2, x\sim 10^{-3}-10^{-4}, A=200$ 
as compared to the DGLAP predictions
(Fig.7). 

\begin{figure}
\vspace*{-.8cm}
\begin{center}     
        \leavevmode 
        \epsfxsize=0.90\hsize 
        \epsfbox{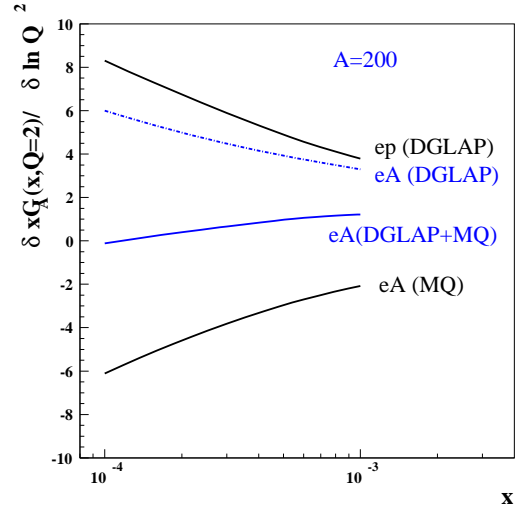} 
    \end{center} 
\label{mqfig}
\vspace*{-1.2cm}
\caption{The scaling violation for the nuclear
 gluon density for A=200
 calculated in the leading twist and with inclusion of the Muller, Qiu 
term.}
\end{figure}

Note however that if
this correction is 
large it certainly cannot be trusted since other effects 
should become important.
  Current models of nonlinear effects try to account for
multinucleon effects, etc. They involve many 
simplifying assumptions.  The overall  trend is an expectation of
large gluon shadowing even if starting from the assumption that at a low
normalization point the leading twist shadowing is small.

Though this approach is seemingly qualitatively different from the 
rest frame approach which we started with and which we will discuss 
in the next subsection these descriptions are in fact dual.

When viewed in the rest frame
deep inelastic scattering corresponds to 
the scattering off a small dipole off the nucleus. The dipole
resolves strong gluon fields at the scale 
 $Q\sim {1\over b}$ and  propagates much longer distances
  $2R_A$  versus  $2r_N$ through the 
gluon fields leading to amplification of nonlinear effects.
Therefore 
 large nonlinear effects are expected based on the space-time picture in 
both reference frames both in the structure of the
 final states and in a very large gluon shadowing which should evolve 
with $Q^2$ in a way substantially different from the DGLAP expectations.

\subsection{The leading twist shadowing for parton densities and hard diffraction 
in electron-proton scattering}

There exists a  deep connection between high-energy
diffraction and phenomenon of nuclear shadowing  \cite{Gribov}.
Qualitatively it  is due to a
 possibility of small momentum transfer
 $-t_{min}=x_{\Pomeron}^2m_N^2$ where  $x_{\Pomeron}=
x_{bj}(1+M^2_{dif}/Q^2)$.
 If  $\sqrt{-t} \le$ ``average nucleon momentum in D(A)''
the amplitudes of diffractive scattering off proton and off 
 neutron interfere. The corresponding double scattering diagram
for the  $\gamma^*D$
 scattering leads
to a diffractive final state.
The  Abramovski$\breve{{\rm i}}$, Gribov, 
 Kancheli theorem  \cite{AGK} then implies that this 
interference for a deuteron
 or a light nucleus:(i)
 increases  the diffractive total 
cross section  by  $\Delta \sigma_{dif}=\sigma_2$, (ii)
 decreases total 
cross section  by  $\Delta \sigma_{tot}=-\sigma_2$, (iii)
decreases
 cross section of inelastic
interactions
with a single  nucleon by  
$\Delta \sigma_{single}= -4\sigma_2$,
(iv) results in simultaneous   inelastic
interactions with two  nucleons with
$\sigma_{double}=2\sigma_2$. The sum of partial contributions 
gives the contribution of the diffraction to the total
cross section: 
$A\sigma_{tot}(eN)-\sigma_{tot}(eA)\equiv 
\Delta \sigma_{tot}=\Delta \sigma_{dif}+\Delta \sigma_{single}
+\sigma_{double}$.

Thus there exists a direct connection between 
$\sigma_{diff}(eN)$
at   $t\sim 0$ and  $\Delta \sigma_{tot}(eA)$
for light nuclei. 
  Neglecting small corrections due to the real part of the 
diffractive amplitude one finds:
\begin{eqnarray}
\Delta \sigma^{ed}_{tot}
={{d\sigma_{diff}(ep)\over dt}
_{\left|t=0\right.}\over 8 \pi R_D^2},
\end{eqnarray}
where  $R_D$ is the deuteron radius. 

As we discussed in section 2  
the QCD factorization theorem is valid for DIS diffraction: 
the  $Q^2$  evolution of
 diffractive structure functions at fixed  $x_{\Pomeron},p_t$ is 
described by DGLAP. 
\begin{figure*}
\vspace*{-1.0cm}
    \begin{center} 
        \leavevmode 
        \epsfxsize=0.9\hsize 
        \epsfbox{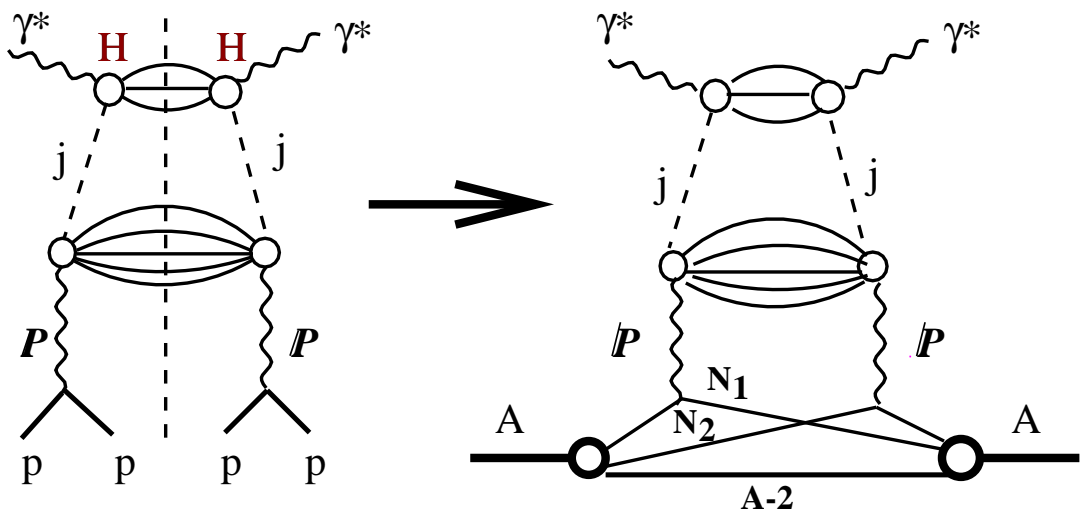}
    \end{center}
\label{shaddia}
\vspace*{-1.0cm}
\caption{Diagrams for hard diffraction in $ep$ scattering
and for the leading twist nuclear shadowing} 
\end{figure*}
By comparing the QCD diagrams for hard diffraction and for nuclear shadowing
due to scattering off two nucleons (Fig.9)
one can prove  \cite{FS98}
 that  in  the low thickness limit the leading twist 
nuclear shadowing is unambiguously expressed through the diffractive
parton densities
 $f_j^D({x\over x_{\Pomeron}},Q^2,x_{\Pomeron},t)$
of $ep$ scattering:
 \begin{eqnarray}
f_{j/A}(x,Q^2)/A  =  f_{j/N}(x,Q^2)\nonumber \\
 -{1 \over 2}
\int d^2b\int_{-\infty}^{\infty}dz_1\int_{z_1}^{\infty} dz_2
\int_x^{x_0} dx_{\Pomeron}\cdot
  \nonumber \\
\cdot f^{D}_{j/N}
\left(\beta, Q^2,x_{\Pomeron},t\right)_{\left|k_t^2=0\right.}
\rho_A(b,z_1)\rho_A(b,z_2)
\cdot
  \nonumber \\
\cos(x_{\Pomeron}m_N(z_1-z_2)),
\label{aeq}
\end{eqnarray}
where  $f_{j/A}(x,Q^2), f_{j/N}(x,Q^2),$ 
 are inclusive 
 parton densities;   $\rho_A(r)$
is the nucleon density in the nucleus.
At not too small $x$ one should add a term related to the 
longitudinal distances comparable to internucleon
 distances in nuclei.
This additional term can be evaluated using information on enhancement 
of
the gluon and valence quark parton densities at $x\sim 0.1$ in
the normalization point as an input. 
It slightly diminishes 
 nuclear shadowing at higher $Q^2$ via the $Q^2$ evolution.
Thus   for light nuclei at small enough $x$ the shadowing effect
for   $1 - f_{j/A}(x,Q^2)/Af_{j/N}(x,Q^2)
\propto \sigma^j_{eff}(x,Q^2)\equiv 
{16\pi d\sigma^{dif}/dt_{\left.\right| t=0 }/ \sigma_{tot}} $.

As we discussed before,  the analyses of the HERA hard diffractive
data indicate that
  $\sigma^g_{eff}(x\le 10^{-3},Q=2) \sim 50
mb, ~~ {\sigma^g_{eff}(x,Q=2)\over \sigma^q_{eff}(x,Q=2)} \sim 3.5$.

\begin{figure*}
\vspace*{-3cm}
\begin{center}     
        \leavevmode 
        \epsfxsize=.80\hsize 
        \epsfbox{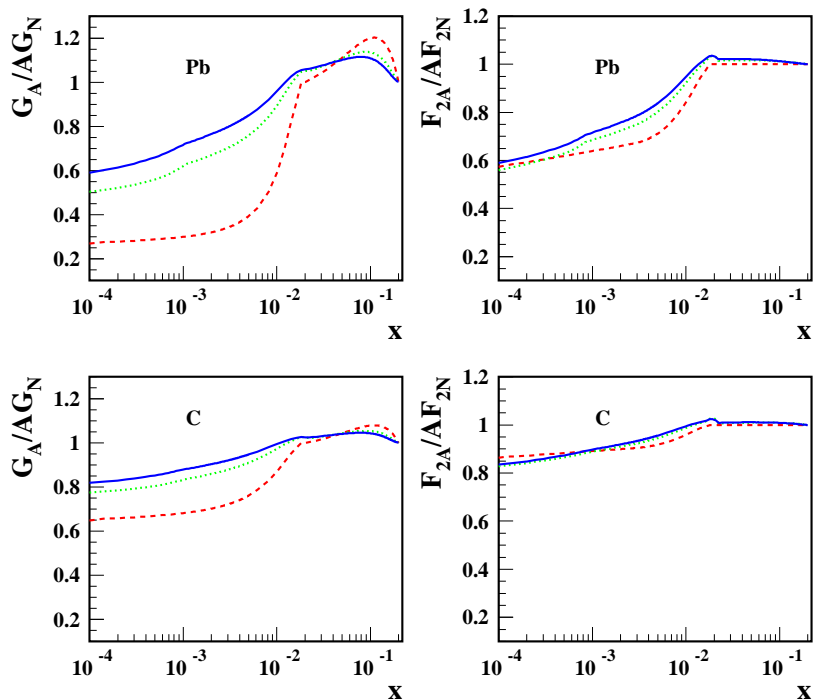}
    \end{center}
\vspace*{-3cm} 
\caption{The  dependence of  $G_A/AG_N$ and $F_{2A}/AF_{2N}$ on  $x$
for Q=2 ( dashed), 5 (dotted),  10 GeV  ( solid)
 curves calculated in
the quasieikonal model using diffractive parton densities of  \cite{ACW}.}
\label{shadfig}
\end{figure*}

 If  one follows indications from  HERA  
that in the  $Q^2\ge 4$ GeV$^2$ range the
leading twist already dominates in  diffraction (this is 
 well established 
 for quarks,  but may not be a good approximation for the 
gluons (see discussion in section 2.2)
one predicts: 
(i) a significant
 shadowing in the quark channel  with small theoretical uncertainty,
(ii)  a much larger (a factor of 3 for
light nuclei)
shadowing in the gluon channel  subject to the question of the
higher twist effects (Fig.\ref{shadfig}). 

 Note than in many models of nuclear shadowing
 which did not incorporate information
about the gluon induced 
diffraction it was expected that the shadowing in the gluon channel
would be smaller than in the quark channel.

The Gribov theory agrees well with the high precision NMC data
if one uses the HERA diffractive data (this involves a certain
extrapolation of the HERA data to smaller  W)  \cite{Orsay}.
A  study of inclusive parton densities at  the  HERA  $x,Q^2$ range 
would (i) establish shadowing in the quark channel, and
check connection between diffraction and shadowing starting from 
nonperturbative domain of  $Q^2 \sim 0$ where shadowing strongly
depends on  $Q^2$, (ii) 
 find out whether interactions in the gluon
channel are superstrong  ($\sim$50 mb) at 
 HERA for intermediate  $Q \sim 2-3
~GeV$, (iii)
study coherent interactions of  $\gamma^*$ with three, four,
... nucleons in the scattering off heavy nuclei. 
Such interactions are sensitive to the fluctuations of the
interaction strength in the quark and gluon channel and check 
the role of weakly interacting components -  this information
 cannot  be obtained from   $ep$ scattering. 
For  A=200 fluctuations may reduce the gluon shadowing 
at  $x=10^{-3} - 10^{-4}$
by a factor up to
 $\sim$ 1.5  \cite{FS98}. Hence
  the study of shadowing for  $ A\ge 40$ would provide a
complementary  information to the case of light nuclei;
(iv) verify connection 
of diffraction and shadowing via investigation
of the $Q^2$ dependence of shadowing between $Q^2=0$ and 4\,GeV$^2$
which should be strong due to a
  strong  $Q^2$ dependence of the diffraction contribution to
 $\sigma_{\gamma^*N}$.

\begin{figure}
\vspace*{-.3cm}
\begin{center}     
        \leavevmode 
        \epsfxsize=0.80\hsize 
        \epsfbox{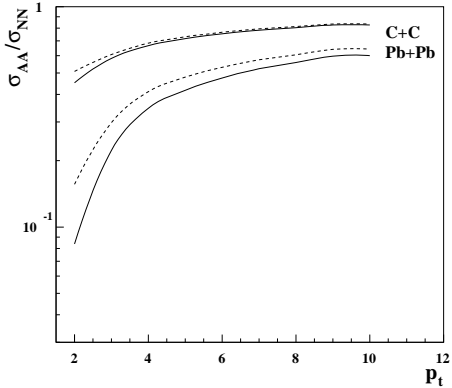}
    \end{center}
\vspace*{-1.1cm} 
\caption{Suppression of the jet production in $AA$ collisions
due to gluon shadowing
at $y=0$ calculated in the quasieikonal and fluctuation models
(solid and dashed curves).}
\label{LHC}
\end{figure}

\subsection{Implications for the heavy ion collisions at LHC}
The production of minijets is often considered as an effective
mechanism of producing high densities in the head on  heavy ion
collisions. For the central rapidities
minijets are produced due to collisions of 
partons with
$x_{jet}={2p_t\over \sqrt{s_{NN}}}$. For heavy ion collisions
at LHC 
$\sqrt{s_{NN}} \sim 4~ TeV$ and
 the gluon-gluon collisions are responsible for
production of most of the  minijets. Therefore the gluon nuclear shadowing
would lead to a reduction of the rate of the jet production
due to the leading twist mechanism by a large
factor up to $p_t\sim 10 GeV/c$, see Fig.\ref{LHC} where we give results
of calculation in the quasieikonal and fluctuation models of shadowing.
The nuclear gluon shadowing leads to a
similar very strong reduction of
the heavy onium production in $pA$ and $AA$ collisions at LHC
energies for $y_{c.m.}\sim 0$ and small $p_t$.
However nuclear shadowing leads to other contributions to the 
jet production not included in the factorization models
- fragmentation of the spectator partons belonging to the dipoles,
decay of the ladders involved in the multiple interactions with 
the dipoles. Hence the overall effect could be a smaller suppression and
additional emission of soft particles over a large rapidity range.
Experimental studies of $eA$ collisions would be of great help in 
quantifying these effects.

\begin{figure}
\vspace*{-.8cm}
\begin{center}     
        \leavevmode 
        \epsfxsize=0.80\hsize 
        \epsfbox{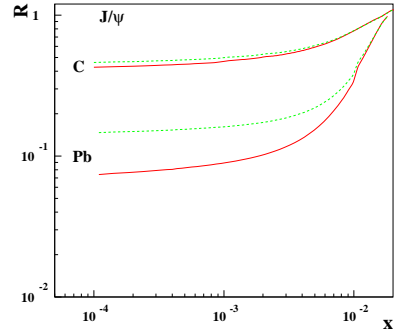}
    \end{center}
\vspace*{-2.0cm} 
\caption{Color opacity effect for the ratio of the coherent production
of  $J/\psi$ from carbon(lead)
and a nucleon normalized to the value of this ratio at  $x=0.02$
(where color transparency is expected to be valid)
calculated in the leading twist approximation 
with (dashed) and
 without ( solid) account of the  fluctuations of 
the interaction strength.}
\label{opacity}
\end{figure}

 \subsection{Study of diffraction with nuclei}
\subsubsection{Exclusive channels: 
 $M\equiv \rho,J/\psi,\pi \pi,...$}
 The QCD factorization theorem for exclusive processes
 \cite{CFS} leads to  \cite{BFGMS} 
\begin{eqnarray}
{{d\sigma\over dt}(\gamma^* A \to M A)\big\vert_{t=0}\over
{d\sigma\over dt}(\gamma^* N \to M N)\big\vert_{t=0}} 
\nonumber \\ =\left [{F^L_A(x,Q) \over F^L_N(x,Q)}\right ]^2
= {G^2_A(x,Q) \over G^2_N(x,Q)}
\end{eqnarray}

Hence in the region $x\ge 0.01$ the color transparency regime of 
weak absorption is expected.
Establishing down to what $Q^2$ this regime would hold would allow to determine
down to what virtualities production of vector mesons 
could be used as a low  $Q^2$ gluonometer.

At smaller $x$
the gluon shadowing would lead to a dramatic 
decrease of the yield of vector meson production at small
$x$ (yields are still large to measure the cross section
accurately) - the color opacity phenomenon - strong absorption of 
small color  dipoles by nuclei at ultra-high energies.
This phenomenon is a qualitative departure from the experience of QED 
where cross section of interaction of small electric dipoles practically does not depend on the incident energy.
The predicted  A-dependence 
for coherent   $J/\psi$ production
is about the same as for photoproduction of  $\rho$-mesons!?

\subsubsection{ Inclusive and semiinclusive channels:}

  Prime objectives of the study of the inclusive
 diffraction is to obtain a direct answer to the question 
 how ``black''
are diffractive interactions. 
We expect an increase
of the rapidity gap 
probability with A. For the heavy nuclei
the fraction of events
where a nucleus remains intact would reach $\sim 30\%$ for the 
quark induced hard diffraction and nearly 50\% for the gluon induced
diffraction  \cite{FSAGK,FS98}.
Also, the rapidity gap probability would depend
on $p_t$ of the jet is a rather peculiar way - 
contrary to naive expectations it 
 would 
increase
between the aligned jet kinematics and the 
$p_t$ range where gluons give dominant contribution.

Overall the study of inelastic diffraction would: 
provide a direct test of the strengths of interaction in the
quark and gluon sector - identify the channels for which interaction is
close to black for heavy nuclei. 
This information will be  complementary to the study of the 
 $A$ dependence of  $G_A/G_N$ and allow to 
clarify  the origin of gluon
component of the ``Pomeron'.

\subsection{Nondiffractive hadron production}
Numerous data on hadron-nucleus scattering at fixed target
energies indicate that the multiplicities of the leading hadrons
$N_A(z)\equiv {1 \over \sigma_{tot}(aA)}{d\sigma(z)^{a+A\to h +X}\over dz}$
decrease with increase of A. Here $z$ is the light-cone fraction of the 
projectile $"a" $ momentum carried by the hadron ``h''.
On the contrary,  the QCD factorization theorem for
the inclusive hadron production in DIS 
implies
that in the case of electron-nucleus scattering no such dependence 
should be present in the  DIS case. 
This indicates that there should be a interesting 
transition from the soft physics dominating in the interactions of
 real photons with nuclei  to the hard physics
in the inclusive hadron production in the DIS kinematics. It would be
 manifested in 
the disappearance
of the A-dependence of the leading spectra at large z:
$N_A(z,Q^2)=N_N(z,Q^2)$, for $z\ge 0.2$, $Q^2 \ge$
 ~few ~GeV$^2$.  
At the same time the transverse spectra are expected to be
 broadened for nuclei ($\Delta p_t^2 \propto A^{1/3}$) due to the QCD analog
the  Landau-Migdal-Pomeranchuk effect, see discussion and references in
 \cite{Baier}.

At small $x$ a new interesting phenomenon should emerge 
 for smaller $z$ due to 
 the presence of diffraction and nuclear shadowing.
Indeed, the diffraction originates from the presence 
in the wave function of $\gamma^*$ of partons with relatively
small
virtualities which screen the color of the leading parton(partons)
 with large  virtuality
and can rescatter elastically from  a target (several target nucleons in
the case of nuclear target). Inelastic interactions of these soft partons
with several nucleons should lead to a
plenty of new revealing phenomena in small $ x$ DIS
$eA$ scattering, which  resemble hadron-nucleus scattering  but with a
shift in rapidity from $y_{max}(current)$
related to the average rapidities of these soft partons. This shift can be 
expressed through the average masses of the hadron states
produced in the diffraction:
\begin{equation}
y_{soft~partons} \sim  y_{max}- \ln (\left<M^2_{dif}\right>/\mu^2),
\end{equation}
where $\mu \sim 1 GeV$ is the soft  scale.
Partons with  these rapidities  will interact in multiple collisions and
loose their energy leading to a dip in the ratio 
$\eta_A(y)\equiv {N_A(y)/N_p(y)}$(Fig.12).
 At the same time these multiple 
interactions should generate a larger multiplicities at smaller rapidities.
Application of the AGK rules indicates that 
for $y \le y_{soft~partons} -\Delta$, where $\Delta = 2-3$
the hadron multiplicity in the case of nuclei will be enhanced
by the factor:
$\eta_A(y)=AF_{2p}(x,Q^2)/F_{2A}(x,Q^2)$.
At the rapidities close to the nuclear rapidities a further increase of 
$\eta_A(y)$ is possible due to formation of hadrons inside the nucleus.
\begin{figure}
\vspace*{-0.2cm}
    \begin{center}
        \leavevmode 
        \epsfxsize=0.9\hsize 
        \epsfbox{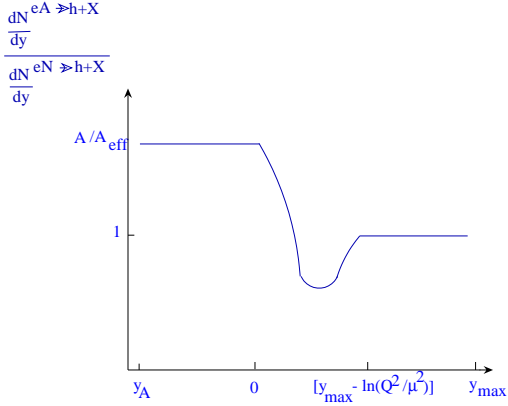}
    \end{center}
\vspace*{-0.5cm}
\caption{A sketch of the A-dependence of the inclusive
hadron production in DIS as a function of rapidity.}
\label{lead}
\end{figure}

One also expects a number of phenomena due to long range correlations 
in rapidity. This includes: (a)  local fluctuations of multiplicity
in the central rapidity region, e.g. the observation of a broader 
distribution of the number of particles per unit rapidity,  due to 
fluctuations of the number of wounded nucleons  \cite{FSAGK}. 
These fluctuations should be larger for the hard processes induced by gluons,
for example the direct photon production of two high $p_t$ dijets.
(b) Correlation of the central multiplicity with the multiplicity of 
neutrons in the forward
neutron detector, etc.

To summarize, the small 
  $x$ ~ $eA$  physics  is one of the  last 
unexplored frontiers of QCD.
A plenty  of new nonlinear QCD phenomena are to be discovered
in these studies: the 
gluon nuclear shadowing, the 
high-energy color transparency, the perturbative color opacity,.. {\it
These are just expected discoveries - what about unexpected ones?}
 Studies of $eA$ collisions would provide 
decisive tests 
of the interface of pQCD and soft physics, allow to
  reach understanding of the space-time
 picture of the high-energy strong interactions. 
 Important applications of these studies include
the measurement of the nonsinglet parton densities in eD
scattering, as well as numerous connections to
 the RHIC and LHC heavy ion physics. 

HERA has produced 
exciting glimpses of new small $x$ strong interaction 
gluon dynamics.
The aim now is to produce a definitive proof and observe
high density hard QCD.

{\bf Acknowledgments:}
One of us (M.S.) would like to thank DESY
 for the hospitality
during the time this work was done. 
We thank 
J. Bartels, J.C. Collins, M. McDermott, A. Mueller,
J. Whitmore
 for discussions of 
 diffractive phenomena.
This work is supported in part by the U.S.
 Department of Energy and BSF.

\end{document}